
%
%
%
%
%
%
\def\ea{{\it et al\/}}
\def\ie{{\it i.e.\ }}
\def\brad{|s_1\rangle}
\def\brai{ \langle s_2|}

%
 \newbox\Ancha
 \def\gros#1{{\setbox\Ancha=\hbox{$#1$}
   \kern-.025em\copy\Ancha\kern-\wd\Ancha
   \kern.05em\copy\Ancha\kern-\wd\Ancha
   \kern-.025em\raise.0433em\box\Ancha}}
%

\input iopppt
\pptstyle
\jl{2}
\noindent To be published in J. Phys. B: At.\ Mol.\ Opt.\ Phys.\ 2002\par

\letter{ A useful form of the  recurrence relation between  relativistic atomic  matrix elements of radial powers}

\author{  R P Mart\'inez-y-Romero,\dag\footnote{\S}{E-mail rodolfo@dirac.fciencias.unam.mx  } H N N\'u\~nez-Y\'epez,\ddag \footnote{$\|$}{E-mail nyhn@xanum.uam.mx}  A L Salas-Brito,$^+$\footnote{\P}{Corresponding author, e-mail: asb@data.net.mx or asb@correo.azc.uam.mx } }[R P Mart\'inez-y-Romero \etal\ ]

\address{\dag Facultad de Ciencias, Universidad Nacional Aut\'onoma de M\'exico, Apartado Postal 50-542, Mexico City, Distrito Federal, CP 04510, Mexico }

\address{\ddag Departamento de F\'isica, Universidad Aut\'onoma Metropolitana- Iztapalapa, Apartado Postal 55-534, Iztapalapa 09340 D F, M\'exico}

\address{$^+$ Laboratorio de Sistemas Din\'amicos, Universidad Aut\'onoma Metropolitana-Azcapotzalco,  Apartado Postal 21-726, C P 04000, Coyoac\'an  D F, M\'exico }

\abs 
Recently obtained recurrence formulae for relativistic hydrogenic radial matrix elements are cast in a simpler and perhaps more useful form.  This is achieved with the help of a new relation between the $r^a$ and the $\beta r^b$ terms ($\beta$ is a $4\times 4$ Dirac matrix and $a,\, b$ are constants) in the atomic matrix elements. \endabs

\pacs{31.30.Jv, 03.65.Pm, 03.20.+i}

The relation of quantum calculations with experimental results is usually made at the level of expectation values or matrix elements of appropriate operators (Moss 1972, Wong and Yeh 1983a,b, De Lange and Raab 1991,  Quiney \etal\ 1997). In atomic or molecular physics this means matrix elements of powers of a radial coordinate between hydrogenic states. This happens both because the hydrogen atom is exactly soluble ---hence very useful in approximate calculation schemes--- and because a power of $r$ can be either a well defined function of interest, like in the London or the Lennard-Jones potentials, or as a term in the multipolar expansion of the electromagnetic field. Given such importance, we have been working in obtaining  recurrence relations between matrix elements of radial functions between hydrogenic  eigenstates but in the  relativistic realm (Mart\'inez-y-Romero \ea\ 2000, 20001). Our purpose here is to report on a new relationship we have recently found between hydrogenic matrix elements of powers of $r$ and of powers of $\beta r$ that can simplify the task of using the recurrence relation previously reported.

By extending to Dirac's relativistic quantum mechanics  a  hypervirial method (N\'u\~nez-Y\'epez \ea\ 1995) that is known to led to the nonrelativistic Blanchard recurrence relation (Blanchard 1974),  it has been possible to obtain the relativistic recurrence formulae (Mart\'inez-y-Romero \ea\ 2000, 20001)

$$ \eqalign{ c_0 \brai r^\lambda \brad =\sum_{i=1}^{3} c_i\brai r^{\lambda -i} \brad + \sum_{i=2}^{3} d_i\brai \beta r^{\lambda -i}\brad, } \eqno(1) $$

\noindent and 

$$  \eqalign{e_0  \brai \beta r^\lambda \brad = &b_0 \brai r^{\lambda}\brad + b_2 \brai r^{\lambda-2}\brad + e_1 \brai \beta r^{\lambda-1}\brad\cr & + e_2 \brai \beta r^{\lambda-2}\brad,} \eqno(2) $$

\noindent  the coefficients $c_i$,  $i=0,\dots 3$ are given by

$$\eqalign{c_0 & = {E^+(E^-)^2\Delta^-_{21}\over E^-\Delta_{21}^- - 4c^2\lambda}, \cr
c_1 & = -{2 Z (E^-)^2 \Delta_{21}^-\over E^-\Delta_{21}^- - 4c^2(\lambda -1)},\cr
c_2 & = c^2{\Delta_{21}^-\Delta_{21}^+\over 4} -\lambda(\lambda -1){E^+\Delta_{21}^-\over E^-\Delta_{21}^- -4c^2\lambda},\cr
c_3& = {-2 Zc^2(\lambda -1)(\lambda -2)\Delta_{21}^-  \over E^- \Delta_{21}^- -4c^2(\lambda -1) },} \eqno(3)$$

\noindent  the coefficients $d_i$, $i=2$ and 3, by

$$ \eqalign{
d_2 & = {c^2\Delta_{21}^-\over 2} \left[(1-\lambda) + {\lambda E^+\Delta_{21}^+\over E^- \Delta_{21}^- -4c^2\lambda}\right],\cr
d_3 & = {Zc^2 (\lambda -1) \Delta_{21}^- \Delta_{21}^+ \over E^- \Delta_{21}^- - 4c^2
(\lambda -1)},}\eqno(4) $$

\noindent and the coefficients $d_i$ and $e_i$ $i=1, 2, 3$  by

$$ \eqalign{b_0=& 4\lambda\left[(E^-)^2 -4 c^4 \right], \cr
             b_2=&c^2(1-\lambda)\left[(\Delta_{21}^{-})^2-4\lambda^2\right], \cr
             e_0=&2E^+[E^-\Delta^-_{21}-4c^2\lambda],\cr
             e_1=&4Z[4c^2\lambda-E^-\Delta^-_{21}],\cr
             e_2=& c^2{\Delta_{21}^+\over 2}[(\Delta_{21}^{-})^2-4\lambda^2].} \eqno(5) $$

\noindent Here we defined $E^\pm\equiv E_2\pm E_1$, $\beta=\hbox{diag}(1, -1)$ is a $2\times 2$ Dirac matrix, the kets $|1\rangle$ and $|2\rangle$   stand for radial eigenstates of the relativistic hydrogen atom, solutions of the Dirac equation $ H_k\psi_k(r) = E_k\psi_k(r),\; k=1,2$, where $k$ is a  label that will serve to distinguish the states appearing in the recurrence relations, $\psi_k(r)$ is a radial bispinor,  and $ \Delta^{\pm}_{21} \equiv  \epsilon_2(2j_2 + 1) \pm \epsilon_1(2j_1 + 1)$. The quantum number $\epsilon$ equals $+1$ when $l=j+1/2$ and equals $-1$ when $l=j-1/2$ where  $j=1/2, 3/2, 5/2, \dots$ is the total angular momentum quantum number, and $l$ is the orbital angular momentum quantum number. It is to be noted that the relativistic states are often labeled by the quantum number  $\kappa$  ---as done in (Drake 1996)--- instead of by $\epsilon$; the relation between these two numbers is    $ \kappa=-\epsilon(j+1/2)$. The possibly complex numbers $\lambda$ are the constant exponents of the radial coordinate $r$.   In the previous equations as in all the paper we use atomic units ($\hbar=m_e=e=1$).

Written  in full, the radial Dirac Hamiltonian $H_k$ and the energy eigenvalues can be written as (Drake 1996, Mart\'inez-y-Romero \ea\ 2000, 20001)

$$ H_k = c\alpha_r\left[p_r - \i\beta{\epsilon_k\over r} \left(j_k+ {1\over 2}\right)\right] +\beta c^2 + V(r),     \eqno(6) $$ 

\noindent where  $p_r=-(\i/r) (1+ r{d\over dr})$, $ \alpha_r $ is a $2\times 2$ antidiagonal matrix having $-1$'s as the only nonzero elements,  $V_k(r)$ is a radial  potential behaving as the fourth component of a 4-vector.   As it must be clear from (6), the  function $\psi_k(r)$ is 

$$ \psi_k(r)\equiv
 {1\over r}\left( \matrix{F_{n_k j_k \epsilon_k}(r)\cr \cr \i G_{n_k j_k \epsilon_k}(r)}\right), \eqno(7) $$

\noindent where $n_k=0, 1, 2, \dots$ is the principal  quantum number,  and $l_k\;=j_k\pm 1/2$, according to whether $l_k$ refers to the big or to the small component of the hydrogenic spinor, so $F_k$ and $G_k$ are, respectively, the   big and  the small component of the radial bispinor with quantum numbers $\epsilon_k$, $j_k$, and $n_k$, corresponding to an energy (Greiner 1991)

$$ E_k\equiv E_{n_k\, j_k}=    c^2 \left (  1+ {Z^2\alpha^2_F \over \left( n_k-j_k-1/2 +\sqrt{(j_k+1/2)^2-Z^2\alpha^2_F }\right)^2} \right)^{-1/2}, \eqno(8)       $$

\noindent where $\alpha_F=1/c\simeq 1/137$ is the fine structure constant. 

The recurrence relations (1) and (2) could be more useful if the matrix elements of $r^\lambda$ were uncoupled from the matrix elements of $\beta r^\lambda$. It is the purpose of this Letter to show  how to achieve such uncoupling [see also the rather cursory discussion in (Ilarraza-Lomel\'i \etal\ 2001)].

To evaluate the relation that is needed to uncouple the recurrence relations, let us first evaluate $H_2\xi+\xi H_1$, that is

$$\eqalign{H_2\xi+\xi H_1=-c^2 \left[2f'{d\over dr}+f'' + {2\over r}f'-\right. &\left. \beta{\Delta_{21}^-\over 2r}{1\over r} f\right] +c^2{\Delta_{21}^-\over 2r}{\Delta_{21}^+\over 2r}f\cr -\i& c\alpha_r \left(f'+\beta {\Delta_{21}^-\over 2r}f \right) 2V,} \eqno(9) $$

\noindent where $\xi(r)\equiv H_2 f(r)-f(r) H_1$, $f(r)$  a radial function, and we indicate with primes $r$-derivatives. Next evaluate 

$$ \eqalign{\i c \left[H_2(-\alpha_r f') +  (-\alpha_r f')H_1\right]= -&c^2\left[2f'{d\over dr}+f'' + {2\over r}f'- \beta{\Delta_{21}^-\over 2r}f' \right]\cr -&2\i c \alpha_rf' V;} \eqno(10)$$

\noindent taking (9) and (10) together, we get

$$  \eqalign{H_2\xi+\xi H_1=  c^2{\Delta_{21}^-\over 2r}{\Delta_{21}^+\over 2r}f +c^2{\Delta_{21}^-\over 2r}\beta \left({f\over r}-f'\right) + \i c \left[H_2(-\alpha_r f')\right.\cr +\left.  (-\alpha_r f')H_1\right]-\i c \alpha_r {\Delta_{21}^-\over r} V(r)\beta f.} \eqno(11) $$

\noindent Combining (11) with

$$E^+ (H_2f-f H_1)= -E^+ \i c \alpha_r \left(f' +{\Delta_{21}^-\over 2 r}\beta f\right), \eqno(12)  $$

\noindent we may get

$$  -(E^+-2V)\i c \alpha_r {\Delta_{21}^-\over 2 r}\beta f =c^2{\Delta_{21}^-\over 2r}{\Delta_{21}^+\over 2r}f +c^2{\Delta_{21}^-\over 2r}\beta \left({f\over r}-f'\right). \eqno(13) $$

\noindent We are almost done with the equation, but we need to eliminate the  term in the left hand side of (13), to this end we have to evaluate the following

 $$ H_2 f+fH_1=-\i c \alpha_r \left[2f{d\over dr}+f' + {2\over r}f+ \beta{\Delta_{21}^+\over 2r} f\right] +2 f(\beta c^2+V). \eqno(14)  $$

\noindent The next steps are analogous to the previous ones, but now using equations (14) and  (10); by further juggling with them, we  get

$$   -(E^+ -2V)\i c \alpha_r {\Delta_{21}^-\over 2 r}\beta f= {\Delta_{21}^-\over 8r^2} \left( \Delta_{21}^- + \Delta_{21}^+\right)(E^+ -2V)\beta f. \eqno(15) $$

\noindent Combining (13) with (15) and taking matrix elements between the states $\langle s_2|$ and $|s_1\rangle$, we finally obtain the fundamental equation for the disentanglement:

$$ g_1 \langle s_2|r^{\lambda-1}|s_1\rangle=j_1 \langle s_2|\beta r^{\lambda-1}|s_1\rangle+j_2 \langle s_2|\beta r^{\lambda-2}|s_1\rangle, \eqno(16) $$

\noindent where we have explicitly used both that $f(r)=r^\lambda$ and the  potential as Coulomb's $V(r)=- Z/r$. The coefficients are given explicitly as

$$ \eqalign{g_1 =& c^2 \left({ \Delta_{21}^- \Delta_{21}^+\over 4}\right),\cr
            j_1 =& {\Delta_{21}^-\over 2 } \left[ {\left( \Delta_{21}^- + \Delta_{21}^+                                                                              \right) \over 4}E^+ + c^2 (\lambda -1) \right], \cr 
            j_2 =& {\Delta_{21}^-\over 4 } Z \left( \Delta_{21}^- + \Delta_{21}^+\right). 
} \eqno (17) $$

  Equation (16) is a recurrence relation that can be found useful for various  tasks but for now  we just want to uncouple  relations (1) and (2). 

 We need first to equate $r^\lambda$ taken from equation (2), to $r^\lambda$ taken from equation (16); in the resulting equation there is yet a term involving $r^{\lambda-2}$ which we eliminate  using again (16). This gives us the following relation between matrix elements  involving $\beta r^\lambda$ terms only,

$$   \eta_0 \langle s_2|\beta r^{\lambda}|s_1\rangle=\eta_1 \langle s_2|\beta r^{\lambda-1}|s_1\rangle+ \eta_2 \langle s_2|\beta r^{\lambda-2}|s_1\rangle+ \eta_3 \langle s_2|\beta r^{\lambda-3}|s_1\rangle, \eqno(18)  $$

\noindent the explicit form of the coefficients is given after equation (19).

To derive the recurrence relation involving matrix elements of $r^\lambda$ terms only, we  get the  term $\beta r^{\lambda-2}$ that arises from equating the term $\beta r^{\lambda-3}$ taken from (1) to the same term taken from (16). From here we also obtain also the terms $\beta r^\lambda$ and $\beta r^{\lambda-1}$ that are to be substituted into (2).  Doing such substitutions we finally obtain the sought after recurrence relation

$$   \eqalign{\nu_0 \langle s_2| r^{\lambda}|s_1\rangle=&\nu_1 \langle s_2| r^{\lambda-1}|s_1\rangle+ \nu_2 \langle s_2| r^{\lambda-2}|s_1\rangle+ \nu_3 \langle s_2| r^{\lambda-3}|s_1\rangle\cr 
&+ \nu_4 \langle s_2| r^{\lambda-4}|s_1\rangle+ \nu_5 \langle s_2| r^{\lambda-5}|s_1\rangle.} \eqno(19)  $$

 \noindent The coefficients in equations (18) and (19) are given by

$$ \eqalign{ 
\eta_0=& {E^+ D \over 2 \lambda F} - {K E^+\over 2 c^2} - {2 \lambda \over \Delta_{21}^+}, \cr 
\eta_1=& Z \left[ {K \over c^2} - {D \over \lambda F} \right], \cr 
\eta_2=& {(\lambda - 1) L \over 2 \lambda F} \left[ {E^+ K \over 4} + {c^2 (\lambda -2) \over \Delta_{21}^+} - {c^2 \Delta_{21}^- \over 4 (\lambda - 1)} \right], \cr 
\eta_3=& { (\lambda - 1) Z K L \over 4 \lambda F}, 
}\eqno (20) $$ 

\noindent and 

$$ \eqalign{ 
\nu_0 =& {2 (E^+)^2 (E^-)^2 J \over c^2 Z (\lambda - 1)}, \qquad \qquad
\nu_1 = -8 E^+ (E^-)^2 {J + 6 c^2 \over c^2 (\lambda - 1)}, \cr 
\nu_2 =& {2 \lambda F \over Z} \left[ {\lambda E^+ \over (\lambda - 1) H} - J - E^+ -{4                     c^2 (\lambda - 2) \over S} \right] - {E^+ D \over Z} \left[ {G \over 2} - {2 c^2                     \over K} \right] \cr 
       & \hskip 1cm - {J \over c^2 Z (\lambda -1)} \left[ 8 Z^2 (E^-)^2 H + {c^2 E^+ (E^-)^2 L \over 2                     H J} \right], \cr 
\nu_3 =& - \left[ D \left( G - {4 c^2 \over K} \right) + 4 E^+ (\lambda - 2) H J + {(E^-                  )^2 L \over (\lambda - 1) } \right], \cr 
\nu_4 =& {c^2 (\lambda -1) L \over 2 Z} \left[ {\lambda E^+ \over (\lambda - 1) H} - J -                     E^+ -{4 c^2 (\lambda - 2) \over S} \right] - 8 Z (\lambda - 2) H J \cr 
       & \hskip 1cm {c^2 \Delta_{21}^+ L \over 2 Z} \left[ {G \over 4} - {c^2 \over K} \right], \qquad\qquad
\nu_5 = c^2 (\lambda - 2) L,
} \eqno (21) $$ 

\noindent where the symbols defined for writing the recurrence relations are:  $M     =  \lambda (\lambda - 1) E^+$, and
$$ \eqalign{  
D     = & \Delta _{21}^- E^- - 4 c^2 \lambda ,\qquad 
J     =  {D + 4 c^2 \over \Delta_{21}^+}, \cr
G     = & {J (2 M - \Delta_{21}^+ D) \over (\lambda -1) D}, \qquad 
F     =  (E^-)^2 - 4 c^4, \cr 
H     = & {D \over D + 4 c^2}, \qquad \qquad
S     =  \Delta_{21}^+ + \Delta_{21}^-, \cr
K     = & {S \over \Delta_{21}^+}, \qquad \qquad
L     =  4 \lambda^2 - (\Delta_{21}^-)^2, \cr 
}\eqno (22) $$

The recurrence relations (16), (18) and (19) hold true if $\omega_1 +\omega_2 +1 +|\lambda| >0$ (Mart\'inez-y-Romero \ea\ 2000, 2001) where the $\omega_a\equiv+\sqrt{(j_a+1/2)^2-Z^2\alpha^2_F},\; a=1,2$ are real numbers. This conditions is basically  that any  integrand goes to zero faster than $1/r$ as $r\to\infty$. As it is easy to realize, the new recursions give no new information in the case of the relativistic Pasternack-Sternheimer (1962) rule ---\ie\ $\Delta_{21}^-=\lambda=0$--- a relation which must then remain as given in our previous contributions (Mart\'inez-y-Romero \ea\ 2000, 2001). The new uncoupled relations should be anyhow useful in relativistic atomic or molecular physics calculations. 

\ack This work  has been partially supported by PAPIIT-UNAM. We acknowledge useful  discussions with A Ilarraza-Lomel\'i and N Vald\'es-Mart\'inez, and the friendly support of Zura,  Mec, Shapka, Rabita, Orejitas, Weiss,  Gorbe,  Bieli  and 
all the gang. \par

\references 
\refjl {Blanchard P 1974} {J. Phys. B: At. Mol. Opt. Phys.} {7} {1993}

\refbk{De Lange O L and Raab R E 1991} { Operator Methods in Quantum Mechanics} {(Oxford: Clarendon)}
\refbk {Drake G W F (Ed) 1996} {Atomic, Molecular and Optical Physics Handbook} {(Woodbury: American Institute of Physics) Ch 22} 

\refbk{Greiner W 1991} {\it Theoretical Physics 3: Relativistic quantum 
mechanics} {(Berlin: Springer)} 
 
\refjl {Ilarraza-Lomel\'i A C,  Vald\'es-Mart\'inez M N, Salas-Brito A L,  Mart\'inez-y-Ro\-mero R P, and  N\'u\~nez-Y\'epez H N 2001} {Int.\ J. Quantum Chem.} {} {contribution submitted to the issue honoring  P. O. L\"owdin}

\refjl {Mart\'inez-y-Romero R P,  N\'u\~nez-Y\'epez H N, Salas-Brito A L 2000} {\it  J.\ Phys.\ B: At.\ Mol.\ Opt.\ Phys.\ } { 33} {L367}

 \refjl {Mart\'inez-y-Romero R P,  N\'u\~nez-Y\'epez H N, Salas-Brito A L 2001} {\it  J.\ Phys.\ B: At.\ Mol.\ Opt.\ Phys.\ } {34}  {1261}

\refbk{Moss R E  1972} {\it Advanced Molecular quantum 
mechanics} {(London: Chapman and Hall)} 

\refjl{N\'u\~nez-Y\'epez H N,  L\'opez-Bonilla J  and Salas-Brito A L 1995} {J. Phys. B: At. Mol. Opt. Phys.} {28} {L525}

\refjl{Pasternack S and Sternheimer R M 1962} {J. Math. Phys.} {3} {1280}

\refjl {Quiney H M, Skaane H and Grant I P 1997} {J. Phys. B: At. Mol. Opt. Phys.} {30} {L829}

\refjl{Wong M K F and Yeh H-Y 1983a} {Phys. Rev. A}  {27} {2300}

\refjl{Wong M K F and Yeh H-Y 1983b} {Phys. Rev. A}  {27} {2305}

\vfill
\eject
\end